\documentstyle[11pt]{article}
\begin{document}
\hoffset=-.5in

\centerline{\Large\bf The Precise Formula in a Sine Function Form}
\centerline{\Large\bf of the Norm of the Amplitude }
\centerline{\Large\bf  and the Necessary and  Sufficient  Phase Condition }
\centerline{\Large\bf for Any Quantum Algorithm with Arbitrary Phase Rotations}
 \footnote{%
The paper was supported by NSFC and  partially by the state key lab. of intelligence technology and system}

\centerline { Dafa Li}

\centerline{ Dept. of Mathematical Sciences} 
\centerline{ National Lab. for
AI at Tsinghua University} 
\centerline{ Tsinghua University Beijing 100084
CHINA} \centerline{email: dli@math.tsinghua.edu.cn} 

\bigskip



Abstract.

In this paper we derived the precise formula in a sine function form of the norm of the
amplitude in the desired state, and by means of the precise formula we
presented the necessary and  sufficient  phase condition for any quantum 
algorithm with arbitrary phase rotations.
 We also showed that the phase condition:  
identical rotation angles
 $\theta =\phi$, is a sufficient but not a necessary phase condition.

\section{Introduction}

\bigskip
Quantum algorithms use two techniques: Fourier transforms \cite{Shor}
 and amplitude amplification \cite{Grover971}\cite{Grover972} \cite{Grover98}.
 Grover's search algorithm is based on the latter above.
The problem addressed by Grover's algorithm is to search 
a desired term(or marked term in \cite{Long1}, or target term in \cite{Grover98})
  in an unordered database
of size $N$. To accomplish this a quantum computer needs $O(\sqrt N )$
queries by using Grover's algorithm \cite{Grover971}.
 In Grover's original version \cite{Grover971} 
the algorithm consists of a sequence of
unitary operations on a pure state, the algorithm is $Q=-I_{0}^{(\pi
)}WI_{\tau }^{(\pi )}W, $where  $W$ is Walsh-Hadamard transformation
and  $I_{x}^{(\pi )}=I-2|x\left\rangle |\right\langle x|,$ which inverts
the amplitude in the state $|x\rangle $; 
here $I_{0}^{(\pi)}$ and $I_{\tau }^{(\pi )}$ invert the amplitudes in the initial and desired basis states $|0 \rangle$ and $|\tau\rangle$, respectively.
To extend his original algorithm Grover
 in \cite{Grover98}  replaced
Walsh-Hadamard transformation  with any quantum
mechanical operation, then obtained the quantum search algorithm $%
Q=-I_{\gamma }^{(\pi )}U^{-1}I_{\tau }^{(\pi )}U, $where $U$ is any unitary
operation and $U^{-1}$ is equal to the adjoint (the complex conjugate of the
transpose) of $U$.  Grover thinks that it leads several new applications and
broadens the scope for implementation. 
To further more generalize Grover's algorithm\cite{Grover98} it allows that
 the amplitudes are rotated by
arbitrary phases, instead of being inverted. 
For example, quantum algorithm $Q=-I_{\gamma
}^{(\theta )}U^{-1}I_{\tau }^{(\phi )}U$, 
$\theta$ and $\phi$ are the rotation angles of the phases of the amplitudes 
in the initial basis state $|\gamma \rangle$
 and in the desired basis state $|\tau\rangle$, respectively.
Recently many authors have been contributing to the general quantum search algorithms 
with any unitary operations and arbitrary phase rotations
\cite{Long1}\cite{LDF}\cite{Hoyer}\cite{Long2}\cite{Long3}\cite{Bilham}.

For the general quantum search algorithms the following problems need to be solved.

1. What is the amplitude in the desired state after k applications of $Q$?

2. What are rotation angles in the initial and the desired basis states 
to reach the desired state from the initial state? 
This problem is called the phase condition. 

3. What is  the optimal number of the iteration steps to find the desired state?

4. Which  of the general algorithms  is the most efficient?

In \cite{LDF} we showed that the amplitude in the
desired state for  any quantum search algorithm which preserves a two-dimensional vector
space  can  be exactly written as a polynomial form in $(\beta \lambda) $.
From the precise formula in a polynomial form in $(\beta\lambda) $
 we obtained some
results in \cite{LDF}. For example we found non-symmetric effects of different rotating
angles and obtained the   approximate formulas of the amplitude of the desired state and the optimal number of iteration steps to find the desired state.
However from the precise formula  it is   not convenient to present a general phase condition. Specially for the algorithms with identical rotation angles Long et al. gave  the approximate formulas of the amplitude in the desired state in \cite{Long1}.
In this paper we will give the precise formulas  in a sine function form of 
the norm of the amplitude in the desired state 
for any quantum search algorithm with arbitrary rotation angles,
which is necessary to present a sufficient and necessary phase condition.

To find the desired state with certainty Long  et al. in %
\cite{Long1} first  presented  a matching condition: 
identical rotation angles $\theta =\phi$. 
Then in \cite{Hoyer} Hoyer gave the phase condition $\tan (\varphi /2)=\tan (\phi /2)(1-a)$. 
In \cite{Bilham} the recursion equation was used to study the
quantum search algorithm, and it concluded that for different rotation angles: 
$\theta \neq \phi$ the algorithm fails to enhance the probability of measuring
a marked state and therefore in order for the algorithm to apply, 
the two rotation angles must be equal, namely, $\theta = \phi$.

In this paper we will study  the general quantum search algorithms 
with any unitary operations and arbitrary phase rotations.
We will give the  phase condition,
 which is  necessary and  sufficient, to find the desired state, 
so we can  thoroughly solve the phase condition  problem presented by Grover in \cite{Grover98}.
 We will also indicate that identical rotation angles $\theta =\phi $, which is the special case of our condition, is  sufficient but not necessary to find the desired state, 
therefore it  contradicts the conclusions obtained in \cite{Long1} and \cite{Bilham}.
Using the precise phase  condition we can construct
 quantum algorithms with arbitrary rotations that succeed with certainty.

This paper is organized as follows.
In section 4 we will derive the precise formula in a sine function form of the norm of the amplitude in the desired state with arbitrary phase rotations.
In section 5  by means of the precise formula in the section 4 we will present
the necessary and  sufficient  phase condition for any quantum algorithm with
arbitrary phase rotations and give the precise optimal number $k_0$ of applications of the algorithm $Q$ to find the desired state. 
In section 6 we will show that identical rotation angles 
$\theta =\phi $ is a sufficient but not a necessary phase
condition to find the desired state.
The  section 7 will introduce 
the $\left| \sin \right| $ property and the periodicity 
and the monotone increasing property in the interval $[0, k_0]$
 of the norm of the amplitude $\left| b_k \right|$ as a function of $k$.
In section 8 and 9 for the algorithms with identical rotation angles $\theta =\phi$ and Grover's algorithm we will give the reduced precise formulas in a sine
function form of the norm of the amplitude and the reduced precise optimal numbers of
iteration steps,  respectively.
In section 10  we will  prove that 
 the optimal number of iteration steps to find the desired state for Grover's algorithm is  less than  the one for the algorithms with arbitrary
identical rotation angles.

\section{The algorithm $Q=-I_{\protect\gamma }U^{-1}I_{\protect\tau }U$,
where $I_{\protect\gamma }=I-2\cos \protect\theta e^{i\protect\theta }|%
\protect\gamma \left\rangle |\right\langle \protect\gamma |$ and $I_{\protect%
\tau }=I-2\cos \protect\phi e^{i\protect\phi }|\protect\tau \left\rangle
|\right\langle \protect\tau |$}

Grover studied the quantum search algorithm \cite{Grover98}: $Q=-I_{\gamma }^{(\pi
)}U^{-1}I_{\tau }^{(\pi )}U,$ where $U$ is any unitary operator and $U^{-1}$
is equal to the adjoint (the complex conjugate of the transpose) of $U$ and
$I_{x}^{(\pi )}=I-2|x\left\rangle |\right\langle x|$
, which inverts the amplitude in the state $|x\rangle$.
 Generally let $%
I_{x}=I-ae^{i\theta }|x\left\rangle |\right\langle x|$. Then $I_{x}$ is
unitary iff $(1-ae^{i\theta })(1-ae^{-i\theta })=1.$ That is, $a=2\cos
\theta $. Then $I_{x}=I-2\cos \theta e^{i\theta }|x\left\rangle
|\right\langle x|.$ Please see \cite{LDF}.  If let $%
I_{x}^{\prime }=I-(ae^{i\theta }+1)|x\left\rangle |\right\langle x|,$ then $%
I_{x}^{\prime }$ is unitary iff $a=\pm 1.$
The case in which $a=-1$ is used in \cite{Long1}.

Let $|\gamma \rangle $ be the initial basis state and $|\tau \rangle $ the desired
basis state. If we apply $U$ to $|\gamma \rangle $, then the amplitude of reaching
state $|\tau \rangle $ is $U_{\tau \gamma }$. That is, $\langle \tau
|U||\gamma \rangle =U_{\tau \gamma }$, and $\langle \gamma |U||\tau \rangle
=U_{\tau \gamma }^{\ast }$, where $U_{\tau \gamma }^{\ast }$ is complex
conjugate of $U_{\tau \gamma }$.

 Let $Q$ be any quantum search algorithm such that

$Q\left( 
\begin{array}{l}
|\gamma \rangle \\ 
U^{-1}|\tau \rangle%
\end{array}%
\right) =M\left( 
\begin{array}{l}
|\gamma \rangle \\ 
U^{-1}|\tau \rangle%
\end{array}%
\right) $, where $M=\left( 
\begin{tabular}{cc}
$\alpha $ & $\beta $ \\ 
$\lambda $ & $\delta  $%
\end{tabular}%
\right) $. That is, $Q$ preserves the vector space spanned by $|\gamma
\rangle $ and $U^{-1}|\tau \rangle .$
After $k$ applications of $Q$ as soon as the state $U^{-1}|\tau \rangle$ is obtained,
then another operation of $U$ will put the state of the quantum computer to 
$|\tau \rangle$, the desired state.

In \cite{LDF}, we introduced the algorithm $Q=-I_{\gamma }U^{-1}I_{\tau }U$,
where $I_{\gamma }=I-2\cos \theta e^{i\theta }|\gamma \left\rangle
|\right\langle \gamma |$ and $I_{\tau }=I-2\cos \phi e^{i\phi }|\tau
\left\rangle |\right\langle \tau |,\alpha =-(1-2\cos \theta e^{i\theta
}+4\cos \theta e^{i\theta }\cos \phi e^{i\phi }|U_{\tau \gamma }|^{2}),$ $%
\beta =2U_{\tau \gamma }\cos \phi e^{i\phi },$ $\lambda =$ $2\cos \theta
e^{i\theta }(1-2\cos \phi e^{i\phi })U_{\tau \gamma }^{\ast },$ $\delta 
=2\cos \phi e^{i\phi }-1$. When $\theta =\phi =0,$ it reduces to Grover's
algorithm. 

Please see \cite{Grover98}\cite{Hoyer}\cite{Long1} to know what  
the matrices $M$ are  like for  Grover's, Long and et al.' and Hoyer's algorithms.

In this paper all discussions and derivations are based on the
algorithm $Q=-I_{\gamma }U^{-1}I_{\tau }U$,
however  the results obtained in this paper
 also hold  for Grover's, Long and et al.'s and Hoyer's algorithms
 and any other quantum search algorithm which preserves a two-dimensional vector
space provided that only $\alpha ,\beta $, $\lambda $ and $%
\delta  $ appear in the results.

\section{ The proof by induction of the precise formula in a
polynomial form in $\protect (\beta \lambda ) $ of the amplitude for
arbitrary phase rotations}

For a quantum search algorithm $Q$ the 
 key problem is what  the amplitude is in the desired state after k applications of $Q$.
In \cite{Long1} 
the approximated formula of the 
amplitude in the desired state was given. In \cite{LDF} the first 
 precise formula  of the amplitude with arbitrary phase rotations was written in 
polynomial form in $\protect (\beta \lambda) $. 
Though we obtained some interesting results in \cite{LDF} by using the precise formula,
it is not convenient to present a general phase condition to find the desired state. 
However the precise formula gave us hints 
in finding the precise formula in a sine function form of the amplitude.   

Let$\ Q|\gamma \rangle =\alpha |\gamma \rangle +\beta (U^{-1}|\tau \rangle
), $ $Q(U^{-1}|\tau \rangle )=\lambda |\gamma \rangle +\delta  (U^{-1}|\tau
\rangle ),$
and $Q^{k}|\gamma \rangle =a_{k}|\gamma \rangle +b_{k}(U^{-1}|\tau \rangle )$%
, where $a_{k}$ and $b_{k}$ are the amplitudes in the state $|\gamma \rangle 
$ and the desired state $U^{-1}|\tau \rangle $, respectively.

In \cite{LDF} we showed  that $a_{k}$ and $b_{k}$ can be exactly written 
as the following polynomial form in $(\beta \lambda ) $, respectively.
Let $[x]$ be the greatest integer which is or less than $x.$

$b_{k}=\beta r_k$, where $r_k=  (c_{k0}+c_{k1}(\beta \lambda )+c_{k2}(\beta \lambda
)^{2}+...+c_{k[(k-1)/2]}(\beta \lambda )^{[(k-1)/2]})$,

\noindent where $c_{kj}=\sum\limits_{n=k-1-2j}^{0}l_{k(k-1-2j-n)}^{(j)}%
\alpha ^{n}\delta  ^{k-1-2j-n}$,and $l_{ki}^{(j)}=\left( 
\begin{tabular}{c}
i+j \\ 
j%
\end{tabular}%
\right) \left( 
\begin{tabular}{c}
k-i-j-1 \\ 
j%
\end{tabular}%
\right) $.

$a_{k}=\alpha ^{k}+d_{k1}(\beta \lambda )+d_{k2}(\beta \lambda
)^{2}+...+d_{k[k/2]}(\beta \lambda )^{[k/2]}$......(5),

\noindent where $d_{kj}=\sum\limits_{n=k-2j}^{0}t_{k(k-2j-n)}^{(j)}\alpha
^{n}\delta  ^{k-2j-n}$, and $t_{ki}^{(j)}=\left( 
\begin{tabular}{c}
i+j-1 \\ 
j-1%
\end{tabular}%
\right) \left( 
\begin{tabular}{c}
k-i-j \\ 
j%
\end{tabular}%
\right) $.Note that $\left( 
\begin{tabular}{c}
n \\ 
0%
\end{tabular}%
\right) =1,$for any $n\geq 0.$
 
However we did not put a strict proof in \cite{LDF}. 
We will put a proof by induction of the conclusion  in appendix 1 of this paper.

\section{For arbitrary phase rotations the precise formula
in a sine function form of the norm of the amplitude \ 
 $\left| b_{k}\right| =\left| \beta \right| \frac{|\sin
k\Delta |}{\sin \Delta }$}

Let$\ Q|\gamma \rangle =\alpha |\gamma \rangle +\beta (U^{-1}|\tau \rangle
), $ $Q(U^{-1}|\tau \rangle )=\lambda |\gamma \rangle +\delta  (U^{-1}|\tau
\rangle ),$
and $Q^{k}|\gamma \rangle =a_{k}|\gamma \rangle +b_{k}(U^{-1}|\tau \rangle )$%
, where $a_{k}$ and $b_{k}$ are the amplitudes in the state $|\gamma \rangle 
$ and the desired state $U^{-1}|\tau \rangle $, respectively. For example, 
$Q^{2}|\gamma \rangle =(\alpha ^{2}+\beta \lambda )|\gamma \rangle $
$+\beta(\alpha +\delta  )(U^{-1}|\tau \rangle ).$ 
Let us  exactly compute $b_{k}$.
From the precise formula in a
polynomial form in $\protect (\beta \lambda ) $ of the amplitude in the section above
  $b_{k}$ $=\beta r_{k}$.  
 Clearly  when $\beta=0$, that is, $\cos \phi =0$, $\left| b_{k}\right| =0$
and the quantum algorithm becomes useless. Therefore we assume that $\beta \neq 0$,
that is,  $\cos\phi \neq 0$ in this paper.

\subsection{A simpler iterated formula of the amplitude  $b_{k}$ }

In \cite{LDF} the
iterated formulas of $a_{k}\ $\ and $b_{k}$ were given as follows. $%
a_{k+1}=(\alpha a_{k}+\lambda b_{k})$ and $b_{k+1}=(\beta a_{k}+\delta 
b_{k}).$ Note that the iterated formula of $b_{k}$ ($a_{k}$) contains the
term $a_{i}$ ($b_{i}$). From the iterated
formula we can derive a simpler iterated formula as follows. From $%
b_{k}=\beta a_{k-1}+\delta  b_{k-1},$we obtain that $a_{k-1}=(b_{k}-\delta 
b_{k-1})/\beta .$Then $b_{k+1}=\beta a_{k}+\delta  b_{k}=\beta (\alpha
a_{k-1}+\lambda b_{k-1})+\delta  b_{k}$

$=\beta \alpha a_{k-1}+\beta \lambda b_{k-1}+\delta  b_{k}=\beta \alpha
(b_{k}-\delta  b_{k-1})/\beta +\beta \lambda b_{k-1}+\delta  b_{k}$

$=(\alpha +\delta  )b_{k}+(\beta \lambda -\alpha \delta  )b_{k-1}.$ Clearly
the formula of $b_{k+1}$ does not contain the term $a_{i},$ therefore it is
simpler than one in \cite{LDF}.

From the section above  $b_{k}$ $=\beta r_{k}$, where $\beta$ does not contain $k$.
So we only need to derive the precise 
formula  in a sine function  form of $r_{k}$. First we derive the iterated formula of $r_k$.

\subsection{The iterated formula of  $r_k$ in the amplitude  $b_{k}$ $=\beta r_{k}$}

From  the precise formula  of the amplitude  $b_{k}$ $=\beta r_{k}$ above, 
after computing  $b_{1}=\beta $ and $b_{2}=\beta (\alpha +\delta  )$, 
then we obtain $r_{1}=1$, $r_{2}=\alpha +\delta  ,$ and $r_{k+1}=(\alpha
+\delta  )r_{k}+(\beta \lambda -\alpha \delta  )r_{k-1}.$

For Grover's algorithm $r_{1}=1$ and $r_{2}=\alpha +1$,$r_{k+1}=(\alpha
+1)r_{k}-r_{k-1}.$

Next let us  derive the precise 
formula of $r_{k}$ using the iterated formula of  $r_k$.

\subsection{ $r_k$ in the amplitude  $b_{k}$ $=\beta r_{k}$ is exactly written as 
$r_{k}=\frac{z_{1}^{k}-z_{2}^{k}}{z_{1}-z_{2}}$, 
where $z_{1}+z_{2}=\alpha +\delta  $, $z_{1}z_{2}=-(\beta \lambda-\alpha \delta  )$}

From the iterated formula that $r_{k+1}=(\alpha +\delta  )r_{k}+(\beta
\lambda -\alpha \delta  )r_{k-1}$ we can derive the precise formula in a sine
function form of $r_{k}$. Let $r_{k+1}=(z_{1}+z_{2})r_{k}-z_{1}z_{2}r_{k-1}$%
, where $z_{1}+z_{2}=\alpha +\delta  $, $z_{1}z_{2}=-(\beta \lambda
-\alpha \delta  )$.
Then we obtain that 
 $(z_{1}-z_{2})r_{k+1}$$=z_{1}^{k+1}-z_{2}^{k+1}$. 
For the  detail derivation please see appendix 2.
From the result 6 in the appendix 3 we know $z_{1}\neq z_{2}$ provided that $%
\cos \phi \neq 0$, then $r_{k+1}=\frac{z_{1}^{k+1}-z_{2}^{k+1}}{z_{1}-z_{2}}$.

\subsection{ $r_k$ in the amplitude  $b_{k}$ $=\beta r_{k}$ is exactly written as 
$|r_{k}|=\frac{|\sin k\Delta |}{\sin \Delta }$}

Let us study the equation $z^{2}-(\alpha +\delta  )z+(\alpha \delta  -\beta
\lambda )=0$ which is the characteristic polynomial of the matrix $M$ which
the algorithm $Q$ corresponds to. Let $|M|$ be the determinant of the matrix 
$M.$ Then $|M|=\alpha \delta  -\beta \lambda =|-I_{\gamma
}^{(x)}U^{-1}I_{\tau }^{(y)}U|=|I_{\gamma }^{(x)}||I_{\tau
}^{(y)}||U^{-1}U|=e^{ix}e^{iy}$, where $x$ and $y$ are the rotation angles.

\ Let $z_{1}$ and $z_{2}$ are the two roots of the equation.
Let $z_{1}=\rho _{1}e^{i\psi _{1}}$ and $z_{2}=\rho _{2}e^{i\psi _{2}},$
where $\rho _{1}>0$ and $\rho _{2}>0.$ From the result 1 in the appendix 3
clearly $\left| z_{1}z_{2}\right| =1,$ then $\rho _{1}\rho _{2}=1$ and $\psi
_{1}+\psi _{2}=2(\theta +\phi ).$ 
Let $z_{1}=\rho e^{i\psi _{1}}$,    where $\rho >0$.
 Then $z_{2}=\frac{1}{\rho }e^{i\psi _{2}}$. 
From the result 2 in the appendix 3 
$z_{1}+z_{2}=\rho e^{i\psi _{1}}+\frac{1}{\rho }e^{i\psi_{2}}$$=2(\cos ((\theta -\phi )-2|U_{\tau \gamma }|^{2}\cos \theta \cos \phi)e^{i(\theta +\phi )}.$

\noindent Then $\rho \cos \psi _{1}+\frac{1}{\rho }\cos \psi _{2}=2(\cos ((\theta -\phi
)-2|U_{\tau \gamma }|^{2}\cos \theta \cos \phi )\cos (\theta +\phi )$,

\noindent and $\rho \sin \psi _{1}+\frac{1}{\rho }\sin \psi _{2}=2(\cos ((\theta -\phi
)-2|U_{\tau \gamma }|^{2}\cos \theta \cos \phi )\sin (\theta +\phi ).$

\noindent Then $\frac{\rho \cos \psi _{1}+\frac{1}{\rho }\cos \psi _{2}}{\rho \sin
\psi _{1}+\frac{1}{\rho }\sin \psi _{2}}=\frac{\cos (\theta +\phi )}{\sin
(\theta +\phi )},$ $\frac{\rho ^{2}\cos \psi _{1}+\cos \psi _{2}}{\rho
^{2}\sin \psi _{1}+\sin \psi _{2}}=\frac{\cos (\theta +\phi )}{\sin (\theta
+\phi )},\rho ^{2}\cos \psi _{1}\sin (\theta +\phi )+\cos \psi _{2}\sin
(\theta +\phi )$

\noindent $=\rho ^{2}\sin \psi _{1}\cos (\theta +\phi )+\sin \psi _{2}\cos (\theta
+\phi ),\rho ^{2}\sin (\psi _{1}-(\theta +\phi ))+\sin (\psi _{2}-(\theta
+\phi ))=0.$ Note that $\psi _{1}+\psi _{2}=2(\theta +\phi )$. Therefore $%
(\rho ^{2}-1)\sin (\psi _{1}-(\theta +\phi ))=0. $

There are two cases. Case 1. In the case $\rho \neq 1.$Thus $\sin (\psi
_{1}-(\theta +\phi ))=0$. From that $\psi _{1}+\psi _{2}=2(\theta +\phi )$
we obtain that $\psi _{1}=\psi _{2}.$Let $z_{1}=\rho e^{i\psi }$. Then $z_{2}=%
\frac{1}{\rho }e^{i\psi }.$Then $z_{1}+z_{2}=(\rho +\frac{1}{\rho })e^{i\psi
}=\alpha +\delta  $, from the results 4 and 5 in the appendix 3 we obtain $%
2<\rho +\frac{1}{\rho }=|\alpha +\delta  |\leq 2.$ Therefore  $\rho \neq 1$
is not possible.

Case 2. In the case $\rho =1.$Let $z_{1}=e^{i\psi _{1}}$ and $z_{2}=e^{i\psi
_{2}}.$ 
Then $r_{k}=\frac{z_{1}^{k}-z_{2}^{k}}{z_{1}-z_{2}}$
$=\frac{e^{ik\psi _{1}}-e^{ik\psi _{2}}}{e^{i\psi _{1}}-e^{i\psi _{2}}}$
$=\frac{\cos k\psi _{1}-\cos k\psi _{2}+i(\sin k\psi _{1}-\sin k\psi _{2})}{%
\cos \psi _{1}-\cos \psi _{2}+i(\sin \psi _{1}-\sin \psi _{2})}$
$=\frac{-2\sin
k\frac{\psi _{1}+\psi _{2}}{2}\sin k\frac{\psi _{1}-\psi _{2}}{2} +2i\cos k%
\frac{\psi _{1}+\psi _{2}}{2}\sin k\frac{\psi _{1}-\psi _{2}}{2} }{-2\sin 
\frac{\psi _{1}+\psi _{2}}{2}\sin \frac{\psi _{1}-\psi _{2}}{2} +2i\cos \frac{%
\psi _{1}+\psi _{2}}{2}\sin \frac{\psi _{1}-\psi _{2}}{2} }$
$=\frac{\sin k\frac{\psi _{1}-\psi _{2}}{2} e^{ik\frac{\psi _{1}+\psi _{2}}{2}%
}}{\sin \frac{\psi _{1}-\psi _{2}}{2} e^{i\frac{\psi _{1}+\psi _{2}}{2}}}$
$=\frac{\sin k\frac{\psi _{1}-\psi _{2}}{2} }{\sin \frac{\psi _{1}-\psi _{2}}{2%
}}e^{i(k-1)\frac{\psi _{1}+\psi _{2}}{2}}$.

From that $\left| z_{1}+z_{2}\right| ^{2}=|\alpha +\delta  |^{2}$
we obtain that $\cos (\psi _{1}-\psi _{2})=2(\cos (\theta -\phi )-2|U_{\tau
\gamma }|^{2}\cos \theta \cos \phi )^{2}-1.$ Since $\cos \phi \neq 0,$ $\psi
_{1}\neq \psi _{2}$, please see the result 6 in the appendix 3$.$ Without
loss of generality, let $\psi _{1}>\psi _{2}$, then $\psi _{1}-\psi
_{2}=\arccos \{2(\cos (\theta -\phi )-2|U_{\tau \gamma }|^{2}\cos \theta
\cos \phi )^{2}-1\}=\arccos (\frac{|\alpha +\delta  |^{2}}{2}-1)$.

Let  $\Delta = \frac{\psi _{1}-\psi _{2}}{2}$. 
Then  $\sin \Delta =\sqrt{1-|\alpha +\delta |^{2}/4} =\sqrt{1-(\cos (\theta
-\phi )-2p^{2}\cos \theta \cos \phi )^{2}}>0 $ 
 Since $\cos \phi \neq 0,$ $\psi_{1}\neq \psi _{2}$.
Therefore $ r_{k}=\frac{\sin k\Delta }{\sin \Delta}e^{i(k-1)(\theta + \phi )}$,
and $|r_{k}|=\frac{|\sin k\Delta |}{\sin \Delta }$. 

\subsection{The precise formula
in a sine function form of the norm of the amplitude\  $\left| b_{k}\right|$}

From that $b_{k}=\beta r_{k}$ we obtain the following versions of $\left|
b_{k}\right| $.

Let $p=|U_{\tau \gamma }|$ in this paper.

Version 1. $\left| b_{k}\right| =\left| \beta \right| \frac{|\sin
k\Delta |}{\sin \Delta }$, where $\left| \beta \right|=2p\left| \cos \phi \right|$. 
Please see the definition of $\Delta $ above.

Version 2. $\left| b_{k}\right| =\left| \beta \right| \frac{|\sin (k(\arccos
(\frac{|\alpha +\delta  |^{2}}{2}-1))/2)|}{\sqrt{1-|\alpha +\delta  |^{2}/4}}$.

Version 3. \ $\left| b_{k}\right| =\left| \beta \right| \frac{|\sin
(k\arcsin \sqrt{1-|\alpha +\delta  |^{2}/4})|}
{\sqrt{1-|\alpha +\delta  |^{2}/4}}$.  

Example 1. $\phi =0$ and $\theta =\pi /2.$ $\left| b_{k}\right| =2p\left|
\sin (k\pi /2)\right| <<1.$Please see the example in \cite{LDF}.
Therefore that $\phi =0$ and $\theta =\pi /2$  can not be used to construct a working  quantum algorithm.

Example 2. $\phi =\pi /6$ and $\theta =\pi /2.$ $\left| b_{k}\right|
=2p\left| \sin (k\pi /3)\right| <<1.$ 
Therefore that $\phi =\pi /6$ and $\theta =\pi /2$ can not be used 
either to construct a working  quantum algorithm.

Comment.
Versions 1 and 2 and 3 of $b_k$  are also the same for Grover's, Long  et al.'s,
Hoyer's algorithms and any other quantum search algorithm which preserves a
two-dimensional vector space.

\section{The necessary and  sufficient  phase condition $\sin \Delta \leq |\beta |$  for arbitrary phase rotations and  the  precise optimal number of iteration steps 
 to find the desired state}

To construct a successful quantum search algorithm we need to know what the rotation angles
are to find the desired state. This is called  the phase condition. 
For arbitrary phase rotations Long et al. first obtained an important result for phase 
condition \cite{Long1}, that is, identical rotation angles $\theta =\phi$. Bilham also obtained the same result using recursion equations  \cite{Bilham}. In \cite{Hoyer} Hoyer gave the phase condition $\tan (\varphi /2)=\tan (\phi /2)(1-a)$. Here we will give a necessary and  sufficient  phase condition for arbitrary phase rotations to find the desired state with certainty. The conditions given by Long et al., Hoyer  and Bilham in \cite{Bilham}\cite{Hoyer} \cite{Long1}  are the special cases of our  phase condition.

\subsection{The necessary and  sufficient  phase condition $\sin \Delta \leq |\beta |$}

An algorithm can search the desired state with certainty, 
that is,there exists a $k$ such that $\left| b_{k}\right| =1,$ 
if and only if $\sin \Delta \leq |\beta |$,
where $\sin \Delta =\sqrt{1-|\alpha +\delta |^{2}/4}$ 
   $ =\sqrt{1-(\cos (\theta -\phi )-2p^{2}\cos \theta \cos \phi )^{2}}.$

Let us derive the conclusion.
If $b_{k}=1$ for some $k$, that is, $|b_{k}|=|\beta |\frac{|\sin k\Delta |}{%
\sin \Delta }=1,$ then $|\sin k\Delta |=\frac{\sin \Delta }{|\beta |}.$
Clearly $\frac{\sin \Delta }{|\beta |}\leq 1$ since $|\sin k\Delta |\leq 1.$
Therefore $\sin \Delta \leq |\beta |$ and the optimal number of iteration
steps to find the desired state with certainty $k_{o}=\frac{1}{\Delta }%
\arcsin \frac{\sin \Delta }{|\beta |}$ such that $|b_{k_{o}}|=1.$Conversely
if $\sin \Delta \leq |\beta |,$ then $\frac{\sin \Delta }{|\beta |}\leq 1.$
Let $k_{o}=\frac{1}{\Delta }\arcsin \frac{\sin \Delta }{|\beta |}$. Clearly $%
k_{o}\Delta =\arcsin \frac{\sin \Delta }{|\beta |}$ and $\sin k_{o}\Delta =%
\frac{\sin \Delta }{|\beta |}$. Therefore $|b_{k_{o}}|=|\beta |\frac{|\sin
k_{o}\Delta |}{\sin \Delta }=1.$
 
Sometimes it is convenient to use $\frac{\sin \Delta }{|\beta |}\leq 1$ 
instead of $\sin \Delta \leq |\beta |$. 
Another version of the  phase condition is 
$\sqrt{1-|\alpha +\delta  |^{2}/4} \leq |\beta |$.
Also $k_{o}$
$=\frac{\arcsin (\sqrt{1-|\alpha +\delta  |^{2}/4}/|\beta |)%
}{\arcsin \sqrt{1-|\alpha +\delta  |^{2}/4}}$.  

The phase condition  also holds  for Long  et
al.'s and Hoyer's algorithms and any other quantum search algorithm which
preserves a two-dimensional vector space.

Example 3. When $\theta =\pi /2,$for any $\phi ,C_{2}=\frac{1}{2p}>>1,$%
therefore $\left| b_{k}\right| <1.$

Therefore that $\theta =\pi /2$ \ can not be used either to construct a working  quantum
algorithm, please see the section 3.3 in \cite{LDF}.

From the  general phase condition we specially get the
following two corollaries whose derivations were put in the appendix 4. It
is convenient to use the two corollaries to check if an algorithm satisfies
 the phase condition.

\subsection{Two corollaries  of the phase condition}

Corollary 1.

Let $\theta $ and $\phi $ be in the same quadrant or $\left| \theta -\phi
\right| <\pi /2$ and $\cos \theta \cos \phi <0$. 
Then $\sin \Delta \leq |\beta |$,  that is,
the algorithm can search the desired state with certainty, if and only if $%
\left| \theta -\phi \right| \leq \arccos (2p^{2}\cos \theta \cos \phi +\sqrt{%
1-4p^{2}\cos ^{2}\phi }).$

Corollary 2.

Let $\theta $ and $\phi $ be in the same quadrant or $\cos \theta \cos \phi
<0$ and $\sin \theta \sin \phi <0$. Then $\sin \Delta > |\beta |$, \ that is, the algorithm
can not search the desired state with certainty,\ if $\left| \sin (\theta
-\phi )\right| >|\beta |  $.

\section{ Identical rotation angles $\protect\theta =\protect\phi $ is a
sufficient but not a necessary phase condition to find the desired state}

For any quantum search algorithm with arbitrary rotation angles what are 
the rotation angles to find the desired state with certainty?
In  \cite{Long1} Long et al. first presented 
identical rotation angles $\theta = \phi$ to find the desired state with certainty.
In \cite{Bilham} Bilham also derived the  phase condition.
In this section we will show that the phase condition:
  identical rotation angles $\theta = \phi$, is sufficient but not a necessary
to find the desired state with certainty.

\subsection{Identical rotation angles $\protect\theta =\protect\phi $ is 
sufficient to find the desired state}

Given $\theta =\phi$ , then $\sin \Delta =2p\left| \cos \phi
\right| \sqrt{1-p^{2}\cos ^{2}\phi },$ 
and clearly  $\frac{\sin \Delta}{|\beta |  }$
$=\sqrt{1-p^{2}\cos ^{2}\phi }%
\leq 1.$ Therefore when $\theta =\phi $ by  the phase condition
 the quantum algorithm $Q$ can search
the desired state with certainty except that $I_{\gamma }=I_{\tau }=I.$

\subsection{Identical rotation angles $\protect\theta =\protect\phi $ is 
not necessary to find the desired state}

We will use the following examples to show the condition $\theta =\phi $ is
not necessary.

Example 4. Assume that $\phi =0.$ Then $\sin \Delta \leq |\beta |$, that is, the
algorithm can search the desired state with certainty,
 if and only if $\left|\sin \theta \right| \leq \frac{2p^{2}}{|1-2p^{2}|}.$

 Let us show the result above holds.
From the  phase condition when $\phi =0$ 
$\frac{\sin \Delta }{|\beta |}=\frac{\sqrt{1-\cos ^{2}\theta (1-2p^{2})^{2}}}{%
2p}.$ Let $\frac{\sin \Delta }{|\beta |} \leq 1$      
Then we obtain $4p^{4}\cos ^{2}\theta \geq (1-4p^{2})\sin ^{2}\theta$,
$\ \sin ^{2}\theta \leq \frac{4p^{4}}{(1-2p^{2})^{2}},$ and $%
\left| \sin \theta \right| \leq \frac{2p^{2}}{|1-2p^{2}|}.$
 
For example, when $p=0.5, \phi =0$ and  $\theta = \frac {\pi}{3}$, 
$\sin \frac {\pi}{3}=\sqrt {3}/2<\frac{2p^{2}}{|1-2p^{2}|}=1$, 
therefore the phase condition is satisfied. In the case $k_0=1$.




Example 5. Assume that $\theta =0.$ It is not hard to show that $\sin \Delta \leq |\beta |$,
that is, the algorithm can search the desired state with certainty,
if and only if $\cos ^{2}\phi \geq 1/(1+4p^{4}).$

From the examples clearly identical rotation angles $\theta =\phi $ is not necessary 
to find the desired state with certainty. 
 Therefore it contradicts the conclusions obtained in \cite{Bilham} and \cite{Long1}.
However identical rotation angles $\theta =\phi $
is a special but an important phase condition.

\section{The $\left| \sin \right| $ property and the periodicity 
and the monotone increasing property in the interval $[0, k_0]$
of the norm of the amplitude $\left| b_k \right|$ as a function of $k$}

For a successful quantum search algorithm $Q$ it is necessary that the norm of the amplitude
in the desired state is amplified after each application of $Q$.
 Therefore to construct a quantum search algorithm we need to study the properties of 
the amplitude in the desired state.
In this section we will show that  any quantum search algorithm with any unitary operations and arbitrary rotation angles is a $|\sin |$ algorithm, here we mean  the norm of the amplitude in the desired state  can be written as $|\sin |$ form after $k$ applications of $Q$. We will also show that though $|b_k|$  periodically changes between 0 and 1
when $k\to \infty$, fortunately as $k$ increases from 0 to the optimal number $k_0$ of iteration steps, $|b_k|$ behaves like $\sin x$ in $[0, \pi/2]$.

\subsection{The $\left| \sin \right| $ property of the norm of the amplitude $\left| b_k \right|$ as a function of $k$}

 From the section 4 above the norm of the amplitude $|b_k|$ in the desired state 
can be written as $|\sin|$ form.
Fix $U_{\tau \gamma}$ and the rotation angles $\theta$ and $\phi$, 
$|b_k|$  is only a $|\sin |$ function of  $k$. 
Therefore it is easy to understand 
why the curves of $\left| b_{k}\right|$ in the Fig. 3 and Fig. 4 on page 30 in \cite{Long1}  look like the ones of $\left| \sin \right| .$ 
In \cite{Long1} Long  et al. could not explain the phenomenon.

\subsection{The periodicity of the norm of the amplitude $\left| b_k \right|$ as a function of $k$}

 Fix  the rotation angles $\theta $, $\phi $ and  $|U_{\tau \gamma }|,$
 $|b_{k}|$ is a periodic function of $k$. 
Let $T$ be the period of $|b_{k}|$ as a $|\sin |$ function of $k$.
Then $T=\frac{\pi}{\Delta}=\pi /\arcsin \sqrt{1-|\alpha +\delta  |^{2}/4}.$
Therefore when  $U_{\tau \gamma}$ and the rotation angles $\theta$ and $\phi$ are fixed, 
then when $k$ tends to infinite,  $|b_k |$, like $|\sin |$, changes between 0 and 1 with the period $T$. 

For the identical rotation angles $\theta =\phi$, the period $T=\pi/\arcsin (2p\left| \cos
\phi \right| \sqrt{1-p^{2}\cos ^{2}\phi })$.
For example, let $p=0.1$ and  $\phi =\pi/4$. Then $T=22.2$. 
In the case the curve of $|b_k|$ looks like  the one in Fig. 3 in \cite{Long1}. 
Let $\phi =9\pi/20$. Then  the curve of $|b_k|$ looks like  the one in Fig. 4 in \cite{Long1}.

For Grover's algorithm  the period $T=\pi /\arcsin (2p\sqrt{1-p^{2}})$.
 Let $p=0.1$ and $0.01$. Then $T=15.7$ and $157$, respectively.

\subsection{The monotone increasing property in the interval $[0, k_0]$,
 where $k_0$ is the optimal number of iteration steps,
of the norm of the amplitude $\left| b_k \right|$ as a function of $k$}

When the phase condition $\sin \Delta \leq |\beta |$ is satisfied, 
the optimal number of iteration steps 
to search the desired state with certainty  $k_{o}=%
\frac{1}{\Delta}\arcsin \frac{\sin \Delta }{|\beta |  } $. 
Since $0<\frac{\sin \Delta }{|\beta |  }%
\leq 1,0<\arcsin \frac{\sin \Delta }{|\beta |  }\leq \pi /2.$
$0<k_{o}\Delta =\arcsin \frac{\sin \Delta}{|\beta |  }\leq \pi /2.$
Therefore when $0 < k\leq k_{o},0<k\Delta \leq k_{o}%
\Delta \leq \pi /2,$ $|\sin k\Delta |=\sin k\Delta ,\left| b_{k}\right| $
$=|\beta | \frac{\sin k\Delta }{\sin \Delta }$.
 Therefore $\left| b_{k}\right| $ as a function of $k$ is strictly
monotone increasing as $k$ increases from 0 to $k_0$.
This result is first reported in this paper.

\section{ In the case  identical rotation angles $\protect\theta =\protect\phi $ the
precise formula in a sine function form of the norm of the amplitude
 and the precise  optimal number of iteration steps}

In \cite{Long1} using many transformations  Long et al. derived
the approximate formulas of the amplitude in the desired state
and the optimal number of iteration steps to find the desired state
for  identical rotation angles.

\subsection{The norm of the amplitude in the desired state}

When $\theta =\phi $ the precise formula  of the norm of the amplitude
 $| b_{k}|$ in the desired state can be reduced.
 In the case let $b_{kl}$ be the amplitude in the desired state. Then we have

Version 1. $\left| b_{kl}\right| =|\beta |  \frac{|\sin
k\Delta |}{\sin \Delta }$, where $\Delta =\frac {\psi_{1}-\psi _{2}}{2}$ and  $\psi
_{1}-\psi _{2}=\arccos \{2(1-2p^{2}\cos ^{2}\phi )^{2}-1\}.$

Remark 1.
Clearly $\Delta$ is small since $p$ is very small.
 Therefore 
$\frac{|\sin k\Delta |}{\sin \Delta }<k$ and 
$\left| b_{kl}\right| < k|\beta |=2kp|\cos \phi |.$

Version 2. $\left| b_{kl}\right| =\frac{\left| \sin k\arcsin (2p\left| \cos
\phi \right| \sqrt{1-p^{2}\cos ^{2}\phi })\right| }{\sqrt{1-p^{2}\cos
^{2}\phi }}$ since $\  \sin \Delta =2p\left| \cos
\phi \right|  \sqrt{1-p^{2}\cos ^{2}\phi }.$

Remark 2.
Let $T$ be the period of $\left| b_{kl}\right|$ as a $|\sin |$ function of $k$.
Then 

\noindent $T=\pi/\arcsin (2p\left| \cos\phi \right| \sqrt{1-p^{2}\cos ^{2}\phi })$.

\subsection{The optimal number  of iteration steps to find the desired state}
Let $k_{ol}$ be the optimal number  of iteration steps to search the desired state with certainty.
Then $k_{ol}=\arcsin \sqrt{1-p^{2}\cos
^{2}\phi }/\arcsin (2p|\cos \phi |  \sqrt{1-p^{2}\cos ^{2}\phi }%
). $

\section{For Grover's Algorithm 
the amplitude in the desired state is exactly written as  
 $\beta \frac{\sin k\xi }{\sin \xi }$ 
and the optimal number of iteration steps is exactly 
 $\arcsin \sqrt{1-p^{2}}/\arcsin 2p\sqrt{1-p^{2}}$}

\subsection{The precise formula in a sine function form
 of the amplitude in the desired state}
 
 For Grover's algorithm $\alpha =1-4|U_{\tau \gamma }|^{2}$, $\beta
=2U_{\tau \gamma }$, $\lambda =-2U_{\tau \gamma }^{\ast }$, $\delta  =1$.
Therefore for Grover's algorithm \ the equation $z^{2}-(\alpha +\delta 
)z+(\alpha \delta  -\beta \lambda )=0$ becomes the equation $z^{2}-(\alpha
+1)z+1=0$ \ which has two conjugate roots since the coefficients are real.
Since the product of the two roots is 1 the norms of the two roots are 1.
Let $z_{1}=e^{i\xi }$ and $z_{2}=e^{-i\xi }.$From $z_{1}+z_{2}=e^{i%
\xi }+e^{-i\xi }=2\cos \xi =\alpha +1,$obtain $\cos \xi =(\alpha
+1)/2=1-2|U_{\tau \gamma }|^{2}=1-2p^2$. Then $z_{1}^{k}-z_{2}^{k}=e^{ik\xi
}-e^{-ik\xi }=2i\sin k\xi ,$ $r_{k}=\frac{z_{1}^{k}-z_{2}^{k}}{%
z_{1}-z_{2}}=\frac{\sin k\xi }{\sin \xi }$, which can also be obtained from
 $r_{k}=$$\frac{\sin k \frac{\psi _{1}-\psi _{2}}{2}}{\sin \frac{\psi _{1}-\psi _{2}}{2}   }   e^{i(k-1)\frac{\psi _{1}+\psi _{2}}{2}}$ 
in the section 4 of this paper
 by letting $\psi _{1}=\xi$ and $\psi _{2}=-\xi$.

 Let $b_{kg}$ be the
amplitude in the desired state for Grover's algorithm. 
Then we have

Version 1.  $b_{kg}=\beta \frac{\sin k\xi }{\sin \xi },$ 
 $|b_{kg}|=|\beta | \frac{|\sin k\xi| }{|\sin \xi|} =2p\frac{|\sin k\xi| }{|\sin \xi|}$,
 where $\cos \xi =1-2p^2$.

Version 2. $\left| b_{kg}\right| =\frac{|\sin k\arccos (1-2p^{2})|}{\sqrt{%
1-p^{2}}}.$

Version 3. $\left| b_{kg}\right| =\frac{|\sin k\arcsin 2p\sqrt{1-p^{2}}|}{%
\sqrt{1-p^{2}}}$.

Remark 1. Fix $U_{\tau \gamma }$, $b_{kg}$ is a $\sin $ function of $k$.

Remark 2. Let $T$ be the period of $|b_{kg}|$ as a $|\sin |$ function of $k$.
 Then $T=\pi /\arcsin (2p\sqrt{1-p^{2}})$. 
$T\rightarrow \infty $ when $p\rightarrow 0.$

Remark 3. $|b_{kg}|=|\beta \frac{\sin k\xi }{\sin \xi }|<2kp$ since $|\frac{\sin k\xi }{\sin \xi }|<k$ when $\xi $ is small.

\subsection{The precise formula of the optimal number of iteration steps
and the derivation of Grover's approximate formula $\frac{\pi}{4p}$}

Let $k_{og}$ be the optimal number of iteration steps to search the desired state with certainty for Grover's algorithm. 
Then  $k_{og}=\arcsin \sqrt{1-p^{2}}/\arcsin
2p\sqrt{1-p^{2}}$ using the version 3 of $|b_{kg}|$ and letting $|b_{kg}| =1$.

In \cite{Grover98} Grover only gave an approximate 
optimal number of applications of Grover's algorithm
 to find the desired state. Let us show how the approximate formula $\frac{\pi}{4p}$
  is derived from our precise formula. 

Clearly  $k_{og}$ can not generate the Taylor's polynomial  at the point $p=0$
since $ \arcsin 2p\sqrt{1-p^{2}}=0$  when $p=0$.
 However $4p/\pi$ is the Taylor's polynomial of degree $1$ generated by $1/k_{og}$ at the point $p=0$. Therefore  $k_{og}$ $\doteq$      $\frac{\pi}{4p}$.

We can also give another explanation for $\frac{\pi}{4p}$.
 Clearly $\pi/2$ is the Taylor's polynomial of degree $0$ 
generated by $\arcsin \sqrt{1-p^{2}}$ at  $p= 0$. 
By using the Taylor's polynomial of degree $1$
 $ \arcsin 2p\sqrt{1-p^{2}}$ $\doteq 2p\sqrt{1-p^{2}} \doteq 2p$.
Therefore $k_{og}$ $\doteq $ $\frac{\pi}{4p}$.
 
Remark 4. $k_{og}\rightarrow \infty $ when $p\rightarrow 0$.

\section{The optimal number of iteration steps for Grover's
algorithm is less than the one for the algorithms 
with arbitrary identical rotation angles $\protect\theta =%
\protect\phi $ when $|U_{\protect\tau \protect\gamma }|$ is fixed}

In the paper \cite{LDF} in the first-order approximate formula of the
amplitude it showed that Grover's algorithm is optimal among algorithms with
arbitrary phase rotations. Now we will strictly show that Grover's algorithm
has the less number of iteration steps than the algorithms with $\theta
=\phi $ when $|U_{\tau \gamma }|$ is fixed.

  $k_{og}$ and $k_{ol}$ are before defined as
the optimal number of iteration steps to search the desired state with certainty
 for Grover's algorithm and for the algorithm with $\theta =\phi $, respectively.
$k_{og}=\arcsin \sqrt{1-p^{2}}%
/\arcsin (2p\sqrt{1-p^{2}})$ and 

\noindent $k_{ol}=\arcsin \sqrt{1-p^{2}\cos ^{2}\phi }/\arcsin
(2p|\cos \phi |  \sqrt{1-p^2\cos ^2\phi }).$
When $\phi=0$ or $\pi$  clearly $k_{ol}$ is reduced to $k_{og}$.

Let show $k_{og}< k_{ol}$  when $0<\phi<\pi/2$ or $\pi /2< \phi < \pi $.

When $p$ is fixed $k_{ol}$ is a
function of $\phi $. And $k_{ol}$ is symmetric about $\phi=\pi/2$.
 Let $(k_{ol})_{\phi }^{^{\prime }}$ be the derivative
of $k_{ol}.$ After computing the derivative we obtain that 
  $(k_{ol})_{\phi }^{^{\prime }}>0$ when $0< \phi < \pi /2 $
therefore  $k_{ol}$ is strictly monotone increasing as $\phi$ increases.
And $k_{ol}$ is strictly monotone decreasing as $\phi$ increases 
when $\pi /2< \phi < \pi $ since  $(k_{ol})_{\phi }^{^{\prime }}<0$.
Therefore always  $k_{og}< k_{ol}$ when $0<\phi<\pi/2$ or $\pi /2< \phi < \pi $. 
$k_{og}= k_{ol}$ only when $\phi=0$ or $\pi$.

 For example $\phi=\pi/3$. 
Then $k_{ol}=\arcsin \sqrt{1-p^{2}/4}/\arcsin (p\sqrt{1-p^{2}/4})$.It is not
hard to see $k_{ol}>$ $k_{og}$  since
$\arcsin \sqrt{1-p^{2}/4}>\arcsin \sqrt{1-p^{2}}$ and $\arcsin (p\sqrt{%
1-p^{2}/4})<\arcsin (2p\sqrt{1-p^{2}}).$

Let $|U_{\tau \gamma }|$ be 0.1. We obtained the following the table 1 using
MATLAB. From the table 1 Clearly  $k_{ol}$ is the least when $\theta =\phi
=0.$Note that when $\theta =\phi =0$ it is just Grover's algorithm.

The table 1

\begin{tabular}{lllllll}
$\theta =\phi =$ & 0 & $\pi /6$ & $\pi /3$ & 2$\pi /3$ & 5$\pi /6$ & $\pi $
\\ 
$k_{ol}=$ & 7 & 8 & 15 & 15 & 8 & 7%
\end{tabular}

\bigskip Acknowledgement. Thank Mr. Cheng Guo for his giving me some idea to
reduce the iterated formula in \cite{LDF} and Xiangrong Li for his computer
experiments using MATLAB at IBM PC.

Appendix 1

{\bf The proof by induction of the precise formula in a
polynomial form in $\protect\beta \lambda $ of the amplitude for
arbitrary phase rotations}

Proof. In \cite{LDF} the iterated formulas of $a_{k}\ $\ and $b_{k}$ were given
as follows. $a_{k+1}=(\alpha a_{k}+\lambda b_{k})$ and $b_{k+1}=(\beta
a_{k}+\delta  b_{k}).$

From the iterated formula of the amplitude by induction hypothesis

$b_{k+1}=\beta a_k +\delta  b_k$$=\beta(\alpha ^{k}+d_{k1}(\beta \lambda
)+d_{k2}(\beta \lambda )^{2}+...+d_{k[k/2]}(\beta \lambda )^{[k/2]})$

$+\delta  \beta (c_{k0}+c_{k1}(\beta \lambda )+c_{k2}(\beta \lambda
)^{2}+...+c_{k[(k-1)/2]}(\beta \lambda )^{[(k-1)/2]})$

$=\beta ((\alpha ^{k}+\delta  c_{k0})+(d_{k1}+\delta  c_{k1})(\beta \lambda
)+(d_{k2}+\delta  c_{k2}) (\beta \lambda)^{2}+....)$

Since $l_{ki}^{(0)}=1$ clearly $c_{k0}=\sum%
\limits_{n=k-1}^{0}l_{k(k-1-n)}^{(0)}\alpha ^{n}\delta ^{k-1-n}$ $=\alpha
^{k-1}+\alpha^{k-2}\delta  +\alpha^{k-3}\delta ^2+...+\delta ^{k-1}$,

and $\alpha ^{k}+\delta  c_{k0}=\alpha ^k +\alpha ^{k-1}\delta +\alpha
^{k-2}\delta  ^2+...+ \delta  ^k=c_{(k+1)0}.$

Next we will show $d_{kj}+\delta  c_{kj}=c_{(k+1)j}$. Note that $%
t_{k0}^{(j)}=\left( 
\begin{tabular}{c}
k-j \\ 
j%
\end{tabular}%
\right) \left( 
\begin{tabular}{c}
j-1 \\ 
j-1%
\end{tabular}%
\right) =l_{(k+1)0}^{(j)}$ and $t_{ki}^{(j)}+l_{k(i-1)}^{(j)}=\left( 
\begin{tabular}{c}
k-j-i \\ 
j%
\end{tabular}%
\right) \left( 
\begin{tabular}{c}
j+i-1 \\ 
j-1%
\end{tabular}%
\right) +\left( 
\begin{tabular}{c}
j+i-1 \\ 
j%
\end{tabular}%
\right) \left( 
\begin{tabular}{c}
k-j-i \\ 
j%
\end{tabular}%
\right) $
$=\left( 
\begin{tabular}{c}
k-j-i \\ 
j%
\end{tabular}%
\right) \left( \left( 
\begin{tabular}{c}
j+i-1 \\ 
j-1%
\end{tabular}%
\right) +\left( 
\begin{tabular}{c}
j+i-1 \\ 
j%
\end{tabular}%
\right) \right) =\left( 
\begin{tabular}{c}
k-j-i \\ 
j%
\end{tabular}%
\right) \left( 
\begin{tabular}{c}
j+i \\ 
j%
\end{tabular}%
\right) =l_{(k+1)i}^{(j)}.$

$d_{kj}+\delta  c_{kj}=t_{k0}^{(j)}\alpha ^{k-2j}+t_{k1}^{(j)}\alpha
^{k-2j-1}\delta  +t_{k2}^{(j)}\alpha ^{k-2j-2}\delta 
^{2}+...+t_{k(k-2j-1)}^{(j)}\alpha \delta  ^{k-2j-1}+t_{k(k-2j)}^{(j)}\delta 
^{k-2j}$

\ \ \ \ \ \ \ \ \ \ \ \ \ \ \ \ \ \ \ \ \ \ \ \ \ $+$ $l_{k0}^{(j)}\alpha
^{k-2j-1}\delta  $ $+$ $l_{k1}^{(j)}\alpha ^{k-2j-2}\delta  ^{2}$\ $%
+...+l_{k(k-2j-2)}^{(j)}\alpha \delta  ^{k-2j-1}$\ $+l_{k(k-2j-1)}^{(j)}\delta 
^{k-2j}$

\ \ \ \ \ \ \ \ \ \ \ \ \  $=t_{k0}^{(j)}\alpha ^{k-2j}$ $+\left(
t_{k1}^{(j)}+l_{k0}^{(j)}\right) \alpha ^{k-2j-1}\delta  +\left(
t_{k2}^{(j)}+l_{k1}^{(j)}\right) \alpha ^{k-2j-2}\delta  ^{2}+$...

$+\left(t_{k(k-2j-1)}^{(j)}+l_{k(k-2j-2)}^{(j)}\right) \alpha \delta 
^{k-2j-1}+\left( t_{k(k-2j)}^{(j)}+l_{k(k-2j-1)}^{(j)}\right)\delta 
^{k-2j}$

$=l_{(k+1)0}^{(j)}\alpha ^{k-2j}+l_{(k+1)1}^{(j)}\alpha ^{k-2j-1}\delta 
+l_{(k+1)2}^{(j)}\alpha ^{k-2j-2}\delta 
^{2}+...$

$+l_{(k+1)(k-2j-1)}^{(j)}\alpha \delta 
^{k-2j-1}+l_{(k+1)(k-2j)}^{(j)}\delta  ^{k-2j}=c_{(k+1)j}.$

When $k=2m+1$, $a_{k}=\alpha ^{k}+d_{k1}(\beta \lambda )+d_{k2}(\beta
\lambda )^{2}+...+d_{km}(\beta \lambda )^{m},$

\ \ \ \ \ \ \ \ \ \ \ \ \ \ \ \ $b_{k}=\beta (c_{k0}+c_{k1}(\beta \lambda
)+c_{k2}(\beta \lambda )^{2}+...+c_{km}(\beta \lambda )^{m}),$

note that $d_{kj}+\delta  c_{kj}=c_{(k+1)j}$, where $1\leq j\leq m,$we
finished the proof of the case.

When $k=2m,a_{k}=\alpha ^{k}+d_{k1}(\beta \lambda )+d_{k2}(\beta \lambda
)^{2}+...+d_{k(m-1)}(\beta \lambda )^{m-1}+d_{km}(\beta \lambda )^{m},$

\ \ \ \ \ \ \ \ \ \ \ $b_{k}=\beta (c_{k0}+c_{k1}(\beta \lambda
)+c_{k2}(\beta \lambda )^{2}+...+c_{k(m-1)}(\beta \lambda )^{m-1}),$

Note that $d_{kj}+\delta  c_{kj}=c_{(k+1)j}$, where $1\leq j\leq m-1,$and
note that $d_{km}=c_{(k+1)m}=1,$we also finished the proof of the case.

As well we can by induction derive the precise formula $a_{k}$ of the
amplitude in the initial state $|\gamma \rangle $ after $k$ applications of $%
Q.$

Therefore the proof is complete.

Appendix 2.

From the iterated formula that $r_{k+1}=(\alpha +\delta  )r_{k}+(\beta
\lambda -\alpha \delta  )r_{k-1}$ we can derive its precise formula in a sine
function form of $b_{k}$. Let $r_{k+1}=(z_{1}+z_{2})r_{k}-z_{1}z_{2}r_{k-1}$%
.....$(1)$, where $z_{1}+z_{2}=\alpha +\delta  $, $z_{1}z_{2}=-(\beta \lambda
-\alpha \delta  )$. Note that $r_{1}=1$ and $r_{2}=\alpha +\delta 
=z_{1}+z_{2} $. Then $r_{k+1}=z_{1}r_{k}+z_{2}r_{k}-z_{1}z_{2}r_{k-1},$thus
we\ obtain 

$r_{k+1}-z_{1}r_{k}=z_{2}r_{k}-z_{1}z_{2}r_{k-1}=z_{2}(r_{k}-z_{1}r_{k-1})$%
....$(2)$

$r_{k+1}-z_{2}r_{k}=z_{1}r_{k}-z_{1}z_{2}r_{k-1}=z_{1}(r_{k}-z_{2}r_{k-1})$%
....$(3)$

From (2) $%
r_{k+1}-z_{1}r_{k}=z_{2}(r_{k}-z_{1}r_{k-1})=z_{2}^{2}(r_{k-1}-z_{1}r_{k-2})=....=z_{2}^{k-1}(r_{2}-z_{1}r_{1}), 
$obtain $r_{k+1}-z_{1}r_{k}=z_{2}^{k-1}(r_{2}-z_{1}r_{1})...(4).$

From (3) $%
r_{k+1}-z_{2}r_{k}=z_{1}(r_{k}-z_{2}r_{k-1})=z_{1}^{2}(r_{k-1}-z_{2}r_{k-2})=....=z_{1}^{k-1}(r_{2}-z_{2}r_{1}), 
$obtain $r_{k+1}-z_{2}r_{k}=z_{1}^{k-1}(r_{2}-z_{2}r_{1})...(5)$

From $z_{1}\times (4)-z_{2}\times (5)$ obtain that $z_{1}r_{k+1}-z_{2}r_{k+1}$
$=z_{1}^{k}(r_{2}-z_{2}r_{1})-z_{2}^{k}(r_{2}-z_{1}r_{1})$
$= z_{1}^{k}(z_{1}+z_{2}-z_{2})-z_{2}^{k}(z_{1}+z_{2}-z_{1})=z_{1}^{k+1}-z_{2}^{k+1}$. 
Note that $r_{2}=z_{1}+z_{2}$ and $r_{1}=1.$

From the result 6 in the Appendix 3 we know $z_{1}\neq z_{2}$ provided that $%
\cos \phi \neq 0$, then $r_{k+1}=\frac{z_{1}^{k+1}-z_{2}^{k+1}}{z_{1}-z_{2}}$%
.

\bigskip

Appendix 3.

Several results

$\alpha =e^{i2\theta }-(e^{i2\theta }+1)(e^{i2\phi }+1)|U_{\tau \gamma
}|^{2},\beta =(e^{i2\phi }+1)U_{\tau \gamma },\lambda =-e^{i2\phi
}(e^{i2\theta }+1)U_{\tau \gamma }^{\ast },\delta  =e^{i2\phi }.$

$\beta \lambda =-e^{i2\phi }(e^{i2\phi }+1)(e^{i2\theta }+1)|U_{\tau \gamma
}|^{2},\alpha \delta  =e^{i2\phi }(e^{i2\theta }-(e^{i2\theta }+1)(e^{i2\phi
}+1)|U_{\tau \gamma }|^{2}),$

Result 1. $\alpha \delta  -\beta \lambda =e^{i2(\theta +\phi )}$. $\ $

Proof.
$(\beta \lambda -\alpha \delta  )=-e^{i2\phi }(e^{i2\phi }+1)(e^{i2\theta
}+1)|U_{\tau \gamma }|^{2}-e^{i2\phi }(e^{i2\theta }-(e^{i2\theta
}+1)(e^{i2\phi }+1)|U_{\tau \gamma }|^{2})$
$=-e^{i2\phi }e^{i2\theta }.$

Let $p=|U_{\tau \gamma }|$ in this paper.

 Result 2. $\alpha +\beta =2(\cos (\theta -\phi )-2p^{2}\cos \theta
\cos \phi )e^{i(\theta +\phi )}$

$\ \ \ \ \ \ \ =2((1-2p^{2})\cos \theta \cos \phi +\sin \theta \sin \phi
)e^{i(\theta +\phi )}.$

Proof.

\ $\alpha +\beta =e^{i2\theta }+e^{i2\phi }-4\cos \theta \cos \phi
e^{i(\theta +\phi )}p^{2}=$

$=(\cos 2\theta +\cos 2\phi -4p^{2}\cos \theta \cos \phi \cos (\theta +\phi
) $

$+i(\sin 2\theta +\sin 2\phi -4p^{2}\cos \theta \cos \phi \sin (\theta +\phi
)$

$=(2\cos (\theta +\phi )\cos (\theta -\phi )-4p^{2}\cos \theta \cos \phi
\cos (\theta +\phi ))$

$+i(2\sin (\theta +\phi )\cos (\theta -\phi )-4p^{2}\cos \theta \cos \phi
\sin (\theta +\phi ))$

$=2(\cos (\theta -\phi )-2p^{2}\cos \theta \cos \phi )(\cos (\theta +\phi
)+i\sin (\theta +\phi ))$

$=2(\cos (\theta -\phi )-2p^{2}\cos \theta \cos \phi )e^{i(\theta +\phi )}$

$=2((1-2p^{2})\cos \theta \cos \phi +\sin \theta \sin \phi )e^{i(\theta
+\phi )}.$

Result 3. $\left| \cos (\theta -\phi )-2p^{2}\cos \theta \cos \phi \right|
\leq 1$ and the equality holds if and only if $\cos \theta =\cos \phi =0.$

 Proof. Note that $\cos (\theta -\phi )-2p^{2}\cos \theta \cos \phi
=(1-2p^{2})\cos \theta \cos \phi +\sin \theta \sin \phi,$ and $-1<1-2p^{2}<1$
whenever $0<p<1.$ When $\cos \theta \cos \phi =0$ clearly $\left|
(1-2p^{2})\cos \theta \cos \phi +\sin \theta \sin \phi \right| \leq 1$. Let
us prove that $\left| (1-2p^{2})\cos \theta \cos \phi +\sin \theta \sin \phi
\right| <1$ whenever $\cos \theta \cos \phi \neq 0$. There are two cases.

Case 1:$\ \cos \theta \cos \phi >0.$

$\ $Case 1.1: $\ 0<p<\sqrt{2}/2.$Then $0<1-2p^{2}<1.$ Then $0<(1-2p^{2})\cos
\theta \cos \phi <\cos \theta \cos \phi ,$
and $\sin \theta \sin \phi <(1-2p^{2})\cos \theta \cos \phi +\sin \theta
\sin \phi <\cos \theta \cos \phi +\sin \theta \sin \phi $
$=\cos (\theta -\phi ).$Therefore $\left| (1-2p^{2})\cos \theta \cos \phi
+\sin \theta \sin \phi \right| <1.$

\ Case 1.2: \ $\sqrt{2}/2\leq p<1.$Then $-1<1-2p^{2}\leq 0.$ Then $-\cos
\theta \cos \phi <(1-2p^{2})\cos \theta \cos \phi \leq 0$, $-\cos (\theta
+\phi )=-\cos \theta \cos \phi +\sin \theta \sin \phi <(1-2p^{2})\cos \theta
\cos \phi +\sin \theta \sin \phi \leq \sin \theta \sin \phi .$ As well $%
\left| (1-2p^{2})\cos \theta \cos \phi +\sin \theta \sin \phi \right| <1.$

Case 2: $\cos \theta \cos \phi <0.$

\ \ Case 2.1: $0<p<\sqrt{2}/2.$Then $0<1-2p^{2}<1.$
$\cos \theta \cos \phi <(1-2p^{2})\cos \theta \cos \phi <0,$ $\cos (\theta
-\phi )<(1-2p^{2})\cos \theta \cos \phi +\sin \theta \sin \phi <\sin \theta
\sin \phi .$
 As well $\left| (1-2p^{2})\cos \theta \cos \phi +\sin \theta \sin
\phi \right| <1.$

\ \ Case 2.2: $\sqrt{2}/2\leq p<1.$Then $-1<1-2p^{2}\leq 0.$
Then $0\leq (1-2p^{2})\cos \theta \cos \phi <-\cos \theta \cos \phi ,$
$\sin \theta \sin \phi \leq (1-2p^{2})\cos \theta \cos \phi +\sin \theta
\sin <-\cos (\theta +\phi ).$

\noindent As well $\left| (1-2p^{2})\cos \theta \cos \phi +\sin \theta \sin \phi
\right| < 1.$

From the cases 1 and 2 when $\cos \theta \cos \phi \neq 0$ clearly $\left|
(1-2p^{2})\cos \theta \cos \phi +\sin \theta \sin \phi \right| <1.$

Now let us to prove the second part.

If $\left| (1-2p^{2})\cos \theta \cos \phi +\sin \theta \sin \phi \right| =1$
then from the above clearly $\cos \theta \cos \phi =0.$
Then $\left| (1-2p^{2})\cos \theta \cos \phi +\sin \theta \sin \phi \right|
=\left| \sin \theta \sin \phi \right| =1.$ Then $\left| \sin \theta |=|\sin
\phi \right| =1,$ that is, $\cos \theta =\cos \phi =0.$

Conversely if $\cos \theta =\cos \phi =0$ then $\left| \sin \theta |=|\sin
\phi \right| =1$. Then $\left| (1-2p^{2})\cos \theta \cos \phi +\sin \theta
\sin \phi \right| =1.$

We finished the proof.

Result 4. $\left| \alpha +\beta \right| =2|\cos (\theta -\phi )-2p^{2}\cos
\theta \cos \phi |\leq 2.$

It is trivial from the results 2 and 3.

Result 5. Assume that $\rho >0.$ Then $\rho +\frac{1}{\rho }\geq 2$, and the
equality \ holds if and only if $\rho =1.$

Result 6. Given that $z_{1}+z_{2}=\alpha +\beta ,z_{1}z_{2}=\alpha \delta 
-\beta \lambda .$ Then $z_{1}\neq z_{2}$\ if $\beta \neq 0$(that is, $\cos
\phi \neq 0)$.

Proof. Assume that $z_{1}=z_{2}$. Let $z_{1}=z_{2}=\rho e^{i\psi }$. From $%
|z_{1}z_{2}|=1$ we obtain $\rho =1.$ From  $2e^{i\psi }=\alpha +\beta $,
we obtain that $\left| \alpha +\beta \right| =2$, that is, $|\cos (\theta -\phi
)-2p^{2}\cos \theta \cos \phi |=1.$From the result 3 in the Appendix 3 it
means that $\cos \theta =\cos \phi =0.$It contradicts that $\cos \phi \neq
0. $ Therefore in the case $z_{1}\neq z_{2}.$

Appendix 4.

From the general phase condition we can derive the following corollaries.
Let $p=|U_{\tau \gamma }|$.

Corollary 1.

Let $\theta $ and $\phi $ be in the same quadrant or $\left| \theta -\phi
\right| <\pi /2$ and $\cos \theta \cos \phi <0$.
 Then $\sin \Delta \leq |\beta |$, that
is, the algorithm can search the desired state with certainty, if and only
if $\left| \theta -\phi \right| \leq \arccos (2p^{2}\cos \theta \cos \phi +%
\sqrt{1-4p^{2}\cos ^{2}\phi }).$

Proof. If $\theta $ and $\phi $ be in the same quadrant then $\cos \theta
\cos \phi >0$ and $\sin \theta \sin \phi >0,$ then $\cos (\theta -\phi
)-2p^{2}\cos \theta \cos \phi =(1-2a^{2})\cos \theta \cos \phi +\sin \theta
\sin \phi >0.$If $\left| \theta -\phi \right| <\pi /2$ and $\cos \theta \cos
\phi <0$ then $\cos (\theta -\phi )>0$ and as well $\cos (\theta -\phi
)-2p^{2}\cos \theta \cos \phi >0.$

$(\Rightarrow ).$ If $\sin \Delta \leq |\beta |$ then $\sqrt{1-(\cos (\theta -\phi
)-2p^{2}\cos \theta \cos \phi )^{2}}\leq |\beta |  $ and $%
1-4p^{2}\cos ^{2}\phi \leq (\cos (\theta -\phi )-2p^{2}\cos \theta \cos \phi
)^{2}.$ In the conditions given $\sqrt{1-4p^{2}\cos ^{2}\phi }\leq \left|
\cos (\theta -\phi )-2p^{2}\cos \theta \cos \phi \right| =\cos (\theta -\phi
)-2p^{2}\cos \theta \cos \phi ,$ and so $\cos (\theta -\phi )\geq \sqrt{%
1-4p^{2}\cos ^{2}\phi }+2p^{2}\cos \theta \cos \phi .$ Therefore $\left|
\theta -\phi \right| \leq \arccos (2p^{2}\cos \theta \cos \phi +\sqrt{%
1-4p^{2}\cos ^{2}\phi })$.

$(\Leftarrow ).$ Clearly $\cos (\theta -\phi )\geq \sqrt{1-4p^{2}\cos
^{2}\phi }+2p^{2}\cos \theta \cos \phi $ and $\cos (\theta -\phi
)-2p^{2}\cos \theta \cos \phi \geq \sqrt{1-4p^{2}\cos ^{2}\phi }$. Given
that $\cos (\theta -\phi )-2p^{2}\cos \theta \cos \phi >0$, obtain $(\cos
(\theta -\phi )-2p^{2}\cos \theta \cos \phi )^{2}\geq 1-4p^{2}\cos ^{2}\phi $%
, and $4p^{2}\cos ^{2}\phi \geq 1-(\cos (\theta -\phi )-2p^{2}\cos \theta
\cos \phi )^{2}\geq 0$ by the result 3 in the appendix 3.
 Therefore $\sin \Delta \leq |\beta |$.
We finished the proof.

\bigskip

In \cite{LDF} we had the result as follows. If $\left| \theta -\phi \right|
<|\beta |  $ then the algorithm $Q$ maybe search the
desired state with certainty. This is the first-order approximate phase
condition. Next we will continue studying what will happen when $\left|
\theta -\phi \right| >|\beta |  .$

Corollary 2.

Let $\theta $ and $\phi $ be in the same quadrant or $\cos \theta \cos \phi
<0$ and $\sin \theta \sin \phi <0$. 
Then $\sin \Delta > |\beta |$, that is, the algorithm
can not search the desired state with certainty,\ if $\left| \sin (\theta
-\phi )\right| >|\beta |  $.

Proof. Note when $\theta $ and $\phi $ are in the same quadrant, $\cos
\theta \cos \phi >0$ and $\sin \theta \sin \phi >0.$

$1-(\cos (\theta -\phi )-2p^{2}\cos \theta \cos \phi )^{2}$

$=\sin ^{2}(\theta -\phi )+4p^{2}\cos (\theta -\phi )\cos \theta \cos \phi
-4p^{4}\cos ^{2}\theta \cos ^{2}\phi $

$=\sin ^{2}(\theta -\phi )+4p^{2}\cos \theta \cos \phi ((1-p^{2})\cos \theta
\cos \phi +\sin \theta \sin \phi )>\sin ^{2}(\theta -\phi )$ in the
conditions given by the corollary. Therefore if $\left| \sin (\theta -\phi
)\right| >|\beta |  $ 
then  $\sin \Delta > |\beta |$. We finished the proof.

\end{document}